\documentclass[twocolumn,prb]{revtex4}

\usepackage{graphicx}
\usepackage{epsfig}
\usepackage{amsmath, amssymb}
\usepackage{verbatim}

\begin{document}

\title{Effective Field Theory and Projective Construction for the \\
$Z_k$ Parafermion Fractional Quantum Hall States}

\author{Maissam Barkeshli}
\author{Xiao-Gang Wen}
\affiliation{Department of Physics, Massachusetts Institute of Technology,
Cambridge, MA 02139, USA }

\begin{abstract}

The projective construction is a powerful approach to deriving the
bulk and edge field theories of non-Abelian fractional quantum Hall
(FQH) states and yields an understanding of non-Abelian FQH states 
in terms of the simpler integer quantum Hall states. Here we show 
how to apply the projective construction to the $Z_k$ parafermion
(Laughlin/Moore-Read/Read-Rezayi) FQH states, which occur at filling
fraction $\nu = k/(kM+2)$. This allows us to derive the bulk low
energy effective field theory for these topological phases, which 
is found to be a Chern-Simons theory at level 1 with a $U(M) \times Sp(2k)$
gauge field. This approach also helps us understand the non-Abelian
quasiholes in terms of holes of the integer quantum Hall states. 

\end{abstract}

\maketitle

\section{Introduction}

Topological order in the quantum Hall liquids is currently the subject
of intense interest because of the possiblity of detecting, for the
first time, excitations that exhibit non-Abelian
statistics.\cite{MR9162,Wnab} On the theoretical side, a primary issue
is how to go beyond some of the known examples of non-Abelian
fractional quantum Hall (FQH) states and to construct and understand
more general non-Abelian FQH phases. 

From the very beginning, two ways to construct and understand
non-Abelian FQH states have been developed.\cite{MR9162,Wnab}  One is
through the use of ideal wave functions and ideal
Hamiltonians.\cite{MR9162,GWW9105} The physical properties of the
constructed FQH states can be deduced using conformal field theory
(CFT).  The other is the projective construction,\cite{Wnab,W9927}
which allows us to derive the bulk effective theory and edge effective
theory for the constructed FQH states.  The physical properties of the
FQH states can be derived from those effective theories.

The $Z_k$ parafermion states at filling fraction $\nu = k/(kM+2)$ were
first studied using the ideal-wavefunction/ideal-Hamiltonian
approach.\cite{MR9162,RR9984} What is the bulk effective theory for
such $Z_k$ parafermion states? When $M=0$, the edge states of the $\nu
= k/2$ $Z_k$ parafermion state are described by the $SU(2)_k$
Kac-Moody (KM) algebra.  Using the correspondence between CFT and
Chern-Simons (CS) theory,\cite{W8951} it was suggested that the bulk
effective theory for the $\nu = k/2$ $Z_k$ parafermion state is the
$SU(2)_k$ CS theory.\cite{FN9804,CF0057} The guessed $SU(2)_k$ CS
theory correctly reproduces the $(k+1)$-fold degeneracy for the $\nu =
k/2$ $Z_k$ parafermion state on a torus.

However, the $SU(2)_k$ CS theory has a serious flaw.  The $SU(2)$
charges in the $SU(2)_k$ KM algebra for the edge states are physical
quantum numbers that can be coupled to external probes, while the
$SU(2)$ charges in the $SU(2)_k$ CS theory are unphysical and cannot
be coupled to external probes without breaking the $SU(2)$ gauge
symmetry. This suggests that the $SU(2)$ in the edge $SU(2)_k$ KM
algebra is not related to the $SU(2)$ in the bulk $SU(2)_k$ CS theory.
This leads us to wonder that the CFT/CS-theory correspondence may not
be the right way to derive the bulk effective theory for generic
non-Abelian states.  In fact, when $M\neq 0$, the edge states for the
$\nu = k/(kM+2)$ $Z_k$ parafermion state are described by
$U(1)_n\otimes Z_k$ CFT, where the $Z_k$ CFT denotes the $Z_k$
parafermion CFT\cite{ZF8515} and $n=k(kM+2)/4$.\cite{CF0057} It is not clear what
is the corresponding bulk effective theory.  Note that the $Z_k$
parafermion CFT can be obtained from the coset construction of the
$SU(2)_k/U(1)$ KM algebra.\cite{GQ8723} This suggests that the bulk
effective theory may be a $SU(2)_k\otimes U(1) \otimes U(1)$ CS
theory.\cite{CF0057}  But a naive treatment of such a CS theory gives
rise to $(k+1)\times$ integer number of degenerate ground states on a torus,
which does not agree with the ground state degeneracy for the  $\nu =
k/(kM+2)$ $Z_k$ parafermion state. We see that the bulk effective theory
for a generic parafermion state is still an unresolved issue.

In this paper, we show how the projective
construction\cite{Wnab,W9927} can be applied to the $Z_k$ parafermion
(Read-Rezayi) states. This leads to a simplified understanding of the
$Z_k$ parafermion states in terms of the integer quantum Hall (IQH)
states and a different way of computing their topological properties.  We
find the bulk effective theory for the $\nu = k/(kM+2)$ $Z_k$
parafermion state to be the 
$[U(M) \times Sp(2k)]_1$ 
CS theory (with a certain choice of electron operators and fermionic
cores for certain quasiparticles).  Such a CS theory correctly
reproduces the ground state degeneracy on a torus.

\section{The projective construction}

The projective construction was explained in detail in \Ref{W9927}.
The idea is to rewrite the electron operator in terms of new fermionic
degrees of freedom:
\begin{equation}
\Psi_e = \sum_{\{\alpha\}} \psi_{\alpha_1} 
...\psi_{\alpha_n} C_{\alpha_1 ...\alpha_n}.
\end{equation}
There are $n$ flavors of fermion fields, $\psi_{\alpha}$, for $\alpha
= 1, \cdots, n$, which carry electromagnetic charge $q_{\alpha}$,
respectively, and which are called ``partons.'' The  $C_{\alpha_1
...\alpha_n}$ are constant coefficients and the sum of the charges of
the partons is equal to the charge of the electron, which we set to 1:
$\sum_\alpha q_\alpha = 1$. The electron operator $\Psi_e$ can be
viewed as the singlet of a group $G$, which is the group of
transformations on the partons that keeps the electron operator
invariant. The theory in terms of electrons can be rewritten in terms
of a theory of partons, provided that we find a way to project the
newly enlarged Hilbert space onto the physical Hilbert space, which is
generated by electron operators.  We can implement this projection at
the Lagrangian level by introducing a gauge field, with gauge group
$G$, which couples to the current and density of the partons. We can
therefore write the Lagrangian as
\begin{equation}
\label{partonL}
\mathcal{L} = i \psi^{\dagger} \partial_0 \psi
+ \frac{1}{2m} \psi^{\dagger} (\partial - i A_i Q)^2 \psi 
+ \text{Tr }(j^\mu a_\mu) + \cdots
\end{equation}
Here, $\psi^{\dagger} = (\psi_{1}^{\dagger}, \cdots, \psi_n^{\dagger})$,
$a$ is a gauge field in the $n\times n$ matrix representation of
the group $G$.  $A$ is the external electromagnetic gauge field and
$Q_{ij} = q_i \delta_{ij}$ is an $n\times n$ matrix with the
electromagnetic charge of each of the partons along the diagonal. The
$\cdots$ respresent additional interaction terms between the partons
and $j^\mu_{ab} = \psi^{\dagger}_a \partial^\mu \psi_b$.
(\ref{partonL}) is simply a convenient rewriting of the theory for the
original electron system in terms of a different set of fluctuating
fields. 

Now we assume that there exists some choice of microscopic interaction
parameters for which the interaction between the partons is such that
the low energy fluctuations of the $a_\mu$ gauge field are weak after
integrating out the partons. This means that the gauge theory that
results from integrating out the partons can be treated perturbatively
about its free Gaussian fixed point.  Since the partons in the absence
of the gauge field form a gapped state $| \Phi_{parton} \rangle$ and
since we can treat the gauge field perturbatively, the
ground state remains to be gapped even after we include the gauge
fluctuations. The ground state wave function is, at least for large
separations, $|z_i - z_j| \gg 1$, of the form
\begin{equation}
\label{wfn}
\Phi(\{z_i\}) = \langle 0 | \prod_{i=1}^N \Psi_e (z_i) |\Phi_{parton} \rangle.
\end{equation}
If we assume that the $i$th parton forms a $\nu = 1$ integer quantum
Hall state, the partons will be gapped and can be integrated out to
obtain an effective action solely in terms of the gauge field. The
action that we obtain is a CS action with gauge group $G$, which should be
expected given that for a system that breaks parity and time-reversal,
the CS term is the most relevant term in the Lagrangian at
long wavelengths. If we ignore the topological properties of the
parton IQH states, then integrating out the partons will yield
\cite{W9927}
\begin{align}
\label{la1}
\mathcal{L} = &\frac{1}{4\pi}  \text{Tr}(a \partial a) + \frac{1}{2\pi} A \text{Tr}(Q \partial a) 
+ \frac{ \text{Tr} (Q^2)}{4\pi} A \partial A + \cdots,
\end{align} 
where $A \partial A = \epsilon^{\mu \nu \lambda} A_\mu \partial_\nu
A_\lambda$ and the $\cdots$ represents higher order terms.  However,
since the partons do not form a trivial gapped state, but rather a
topologically non-trivial one, eqn. (\ref{la1}) can only describe
ground state properties of the phase. It can be expected to reproduce
the correct result for the ground state degeneracy on genus $g$
surfaces, for instance, and the correct fusion rules for the
non-Abelian excitations, but it cannot be expected to produce all of
the correct quantum numbers for the quasiparticle excitations, such as
the quasiparticle spin,\footnote{The quasiparticle spin is also
sometimes referred to as the ``twist'' and is equal to the scaling
dimension modulo 1 of the corresponding quasiparticle operator on the edge.}
unless the partons are treated more carefully. This can be done in two
ways.  One way is to not integrate out the partons and to use
(\ref{partonL}), taking into account a Chern-Simons term for $a_\mu$
that emerges as we renormalize to low energies. As will be discussed
in more detail in Section \ref{qps}, the quasiparticles will
correspond to holes in the parton IQH states which become non-Abelian
as a result of the coupling to the non-Abelian Chern-Simons gauge
field.  The other way is to use the pure gauge theory in (\ref{la1})
and to put in by hand a fermionic core for quasiparticles that lie in
certain ``odd'' representations of $G$. Some quasiparticles correspond
to an odd number of holes in the parton IQH states and the fermionic
character of these odd number of holes should be taken into account. 

Let us now turn to the edge theory. Before the introduction of the
gauge field, the edge theory is the edge theory for $n$ free fermions
forming an integer quantum Hall state. If each parton forms a $\nu =
1$ IQH state, then the edge theory would be a CFT describing $n$
chiral free fermions, which we will denote as $U(1)^n$. After
projection, the edge theory is described by a $U(1)^n/G$ coset theory
that we will understand in some more detail when we specialize to the
$Z_k$ parafermion states.

To be more precise, the edge theory should be understood in the
following way. The electron creation and annihilation operators,
$\Psi_e$ and $\Psi_e^{\dagger}$, generate an operator algebra that we
refer to as the electron operator algebra.  Such electron operator
algebra can be embedded in the $U(1)^n/G$ coset theory.  The
topologically distinct quasiparticles are then labelled by different
representations of this electron operator algebra. In some cases, the
electron operator algebra coincides with some well-known algebra.  For
the bosonic $Z_k$ parafermion states at $\nu = k/2$, for instance, the
electron operator algebra is the same as the $SU(2)_k$ KM
algebra, for which the representation theory is well-known. 


\section{Effective theory of parafermion states}

Now let us apply the projective construction to obtain the $Z_k$
parafermion states. A crucial result for the projective construction
is that the $\nu = 1$ FQH wave function coincides with the correlation
function of free fermions in a 1+1d CFT:
\begin{align}
\label{v1corr}
\prod_{i<j} (z_i - z_j) = 
\lim_{z_\infty \rightarrow \infty} z_{\infty}^{2h_N} \langle 
e^{-iN\phi(z_{\infty})} \psi(z_1) \cdots \psi(z_N) \rangle,
\end{align}
where $\psi(z)$ is a free complex chiral fermion, and $\partial \phi = \psi^{\dagger}
\psi$ is the fermion current. The operator product expansions for $\psi(z)$ satisfy
$\psi^{\dagger}(z) \psi(w) \sim \frac{1}{z - w}$ and 
$\psi(z) \psi(w) \sim (z-w) \psi \partial \psi (w)$.
Eqn. (\ref{v1corr}) implies that the wave function
(\ref{wfn}) can also be expressed as a correlation function in a 1+1d
CFT:\cite{W9927}
\begin{align}
\label{wfnCorrel}
\Phi(\{z_i\}) = \lim_{z_\infty \rightarrow \infty} z_{\infty}^{2h_N} \langle e^{-iN\phi(z_{\infty})} \prod_i \Psi_e(z_i) \rangle,
\end{align}
where the partons $\psi_i(z)$ are now interpreted as free fermions in
a 1+1d CFT.

The $Z_k$ parafermion FQH wave functions are constructed as
correlation functions of a certain CFT:
\begin{equation}
\Phi_{Z_k} = \lim_{z_\infty \rightarrow \infty} z_{\infty}^{2h_{N}} \langle e^{-i N \phi(z_\infty)} V_e(z_1) \cdots V_e(z_N) \rangle,
\end{equation}
where $V_e = \psi_1 e^{i\sqrt{1/\nu} \phi}$. $\psi_1$ is a
simple-current operator in the $Z_k$ parafermion CFT of Zamalodchikov
and Fateev\cite{ZF8515} and $\phi$ is a free scalar boson.  These wave functions
exist for $\nu = \frac{k}{kM + 2}$; for $M = 0$, the electron operator
$V_e = \psi_1 e^{i\sqrt{2/k} \phi} \equiv J^+$ and $V_e^{\dagger} =
\psi_1^{\dagger} e^{-i\sqrt{2/k}\phi} \equiv J^-$ generate the $SU(2)_k$
KM algebra: 
\begin{align} J^a(z) J^b(0) \sim \frac{k \delta_{ab}}{z^2} +
\frac{if_{abc} J^c(0)}{z} + \cdots,
\end{align}
where $a,b = 1, 2, 3$ and $J^{\pm} = J^1 \pm i J^2$.  This means that
any electron operator that satisfies the $SU(2)_k$ current algebra
will yield the same wave function. The crucial result for the
projective construction approach to the $Z_k$ parafermion states is
that if we take the electron operator to be 
\begin{equation}
\Psi_{e;k} = \sum_{a=1}^{k} \psi_{2a -1}\psi_{2a},
\end{equation}
then it is easy to verify that $\Psi_{e;k}$ and $\Psi_{e;k}^{\dagger}$ also
satisfy the $SU(2)_k$ current algebra and therefore the wave function
(\ref{wfnCorrel}) is the $Z_k$ parafermion wave function. It follows
that the $Z_k$ parafermion states at $\nu = \frac{k}{kM+2}$, for
general $M$, are reproduced in the projective construction for the
following choice of electron operator 
\begin{equation} 
\Psi_e^{(k;M)} = \psi_{2k+1} \cdots \psi_{2k+M} \sum_{a=1}^{k} \psi_{2a -1}\psi_{2a},
\end{equation}
because including the additional operators $\psi_{2k+1}, \cdots,
\psi_{2k+M}$, each of which is in a $\nu = 1$ IQH state, has the effect of
multiplying $\Phi_{Z_k}$ by the Jastrow factor $\prod_{i<j} (z_i -
z_j)^M$. 

In the case $M = 0$, the electron operator can be written as
$\Psi_e^{(k;0)} = \psi^T A \psi$, where $\psi^T = (\psi_1, \cdots,
\psi_{2k})$ and $A = \left( \begin{matrix} 0 & -\mathbb{I} \\
\mathbb{I} & 0 \end{matrix} \right)$.\footnote{To write the electron
operator this way, we have renumbered the partons.} $\mathbb{I}$ is
the $k \times k$ identity matrix. The group of transformations on the
partons that leaves the electron operator invariant is simply the
group of $2k \times 2k$ matrices that keeps invariant the
antisymmetric matrix $A$. In this case, this group is the fundamental
representation of $Sp(2k)$. Note that $Sp(2) = SU(2)$ and $Sp(4) =
SO(5)$.  Thus, we expect the edge theory to be $U(1)^{2k}/Sp(2k)_1$,
and the bulk CS theory to be $Sp(2k)_{1}$, as
described in the previous section.  For general $M$, the edge
theory becomes $U(1)^{2k+M}/[U(M)\times Sp(2k)]_1$ and the bulk
effective theory is a $[U(M) \times Sp(2k)]_1$ Chern-Simons theory.

\section{Ground state degeneracy from effective CS theory}

As a first check that this CS theory reproduces the correct
topological properties of the $Z_k$ parafermion states, we calculate
the ground state degeneracy on a torus.  This can be done explicitly
using the methods of \Ref{W9927,WZ9817}; for $M = 0$, the result is 
$k + 1$, which coincides with the torus degeneracy of the $M = 0$ $Z_k$
parafermion states. In Appendix \ref{gsdeg} we outline in more detail
the calculation in the case $M = 1$, for which we find the ground state
degeneracy on a torus to be $(k+1)(k+2)/2$, which also agrees with
known results for the $Z_k$ parafermion states. 

The case $M = 1$ reveals a crucial point. In this case, 
we have $[U(1) \times Sp(2k)]_1$ CS theory. Naively, we would think that the extra
$U(1)_1$ part is trivial and does not contribute to the ground state
degeneracy or the fusion rules, and again we might expect a ground
state degeneracy of $k + 1$, but this is incorrect. The reason for
this is that usually when we specify the gauge group and the level for
CS theory, there is a standard interpretation of what the
large gauge transformations are on higher genus surfaces, but this
standard prescription may be inapplicable.  Instead, the large gauge
transformations are specified by the choice of electron operator.
In particular, for odd $k$, the extra factor $(k+2)/2$ is half-integer,
which highlights the fact that the $U(1)$ and $Sp(2k)$ parts are married
together in a non-trivial way. 

In the $M = 0$ case, the standard interpretation of the allowed gauge
transformations for the $Sp(2k)_1$ CS theory is correct, and
we can follow the standard prescription for deriving topological
properties of CS theories at level $k$ with a simple Lie
group $G$. In these cases, the ground state degeneracy is given by the
number of integrable representations of the affine Lie algebra
$\hat{g}_k$, where $g$ is the Lie algebra of $G$. The quasiparticles
are in one-to-one correspondence with the integrable representations
of $\hat{g}_k$, and their fusion rules are identical as well. In the
case of the $M = 0$ $Z_k$ parafermion states, it is already known that
the different quasiparticles correspond to the different integrable
representations of the $SU(2)_k$ KM algebra, and the fusion
rules are the same as the fusion rules of the $SU(2)_k$
representations.  In fact, $Sp(2k)_1$ and $SU(2)_k$ have the same
number of primary fields and the same fusion rules, and so the
$Sp(2k)_1$ CS theory has the same fusion rules as the $Z_k$
parafermion states and the same ground state degeneracies on high
genus Riemann surfaces.  The equivalence of the fusion rules for the
representations $Sp(2k)_1$ and $SU(2)_k$ current algebra is a special
case of a more general ``level-rank'' duality between $Sp(2k)_n$ and
$Sp(2n)_k$,\cite{C9191} and is also related to the fact that the edge
theory for the $M = 0$ $Z_k$ parafermion states can be described
either by the $U(1)^{2k}/Sp(2k)_1$ coset theory or, equivalently, by
the $SU(2)_k$ Wess-Zumino-Witten model. For a more detailed discussion, see
Appendix \ref{lrd}.

\section{Quasiparticles from the projective construction}
\label{qps}

We can understand the non-Abelian quasiparticles of the $Z_k$ FQH
states as holes in the parton integer quantum Hall states. \footnote{
This is related to the observation in \Ref{CG0199} that the $Z_k$ parafermion quasihole 
wave functions can be obtained by symmetrizing or anti-symmetrizing the quasihole wave functions
of a generalized (331) state.}
After projection, these holes become the non-Abelian quasiparticles and we
can analyze these quasiparticles using either the bulk CS
theory or through the edge theory/bulk wave function, all of which we
obtained from the projective construction. The easiest way
to analyze the quasiparticles is through the latter approach, which we
describe first.  The fundamental quasihole is the one with a single
hole in one of the parton IQH states. We expect the wave function
for this excited state to be, as a function of the quasiparticle 
coordinate $\eta$ and the electron coordinates $\{z_i\}$,
\begin{align} 
\Phi_\gamma(\eta&; \{z_i \}) \sim \langle 0 | \prod_i \Psi_e(z_i) \psi_1^{\dagger}(z_{\infty})
\psi_1(\eta) | \Phi_{parton} \rangle \nonumber \\ 
&\sim \langle e^{-i(N+q_1) \phi(z_\infty)} \prod_i \Psi_e(z_i) \psi_1(\eta) \rangle.
\end{align}
More general quasiparticles should be related to operators of the form
$\psi_i \psi_j \psi_k \cdots$. To see whether these operators really
correspond to the non-Abelian quasiparticles of the $Z_k$ parafermion
states, we can study their pattern of zeros.\cite{WW0809,BW0932} The
pattern of zeros is a quantitative characterization of quasiparticles
in the FQH states. In general, it may not be a
complete one-to-one labelling of the quasiparticles, but in the case
of the $Z_k$ parafermion states, it is; one way to see this from the
projective construction approach is to compute the ground state
degeneracy on the torus from the projective construction, which yields
the number of topologically distinct quasiparticles, and then to
observe that the number of operators with distinct pattern of zeros is
the same as the number of distinct quasiparticles. 

The pattern of zeros $\{l_{\ga;a}\}$ is defined as follows.\cite{BW0932} Let
$V_{\ga}$ denote the quasiparticle operator, and let $V_{\ga;a} =
\Psi_e^a V_{\ga}$. Then, \begin{equation} \Psi_e(z) V_{\ga;a}(w) \sim
(z-w)^{l_{\ga;a+1}} V_{\ga;a+1} + \cdots,
\end{equation}
where $\cdots$ represent terms higher order in powers of $(z-w)$.
From $\{l_{\ga;a}\}$ we construct the occupation number sequence
$\{n_{\ga;l}\}$ by defining $n_{\ga;l}$ to be the number of $a$ for
which $l_{\ga;a} = l$. The occupation number sequences $n_{\ga;l}$ are
periodic for large $l$ and topologically distinct quasiparticles will
have occupation numbers with distinct unit cells for large $l$. In
Table \ref{pozTable}, we have listed pattern of zeros for some
of the operators of the form $\psi_i \psi_j \cdots$.  We see that 
they coincide exactly with the known quasiparticle pattern of zeros 
in the $Z_k$ parafermion states, indicating that these operators do 
indeed correspond to the quasiparticle operators of the $Z_k$ 
parafermion states.  Note that two sets of operators 
correspond to topologically equivalent quasiparticles if either they 
can be related to each other by a gauge transformation or
by the electron operator. In Table \ref{pozTable}, some of the gauge
equivalences are indicated, using the symbol $\sim$. There are also
various operators that are not simply gauge equivalent but that also differ by
electron operators. For example, in the $Z_3$ states for $M = 0$, 
the operators $\psi_1$ and $\psi_1 \psi_2 \psi_j$ are topologically
equivalent quasiparticle operators; for the $Z_2$ states at $M = 0$,
$\psi_i$ and $\psi_i \psi_j \psi_k$ are also topologically equivalent,
\it etc. \rm

\begin{table} 
\begin{tabular}{|c c c|} \hline
\multicolumn{3}{|c|}{$Z_2$ states, $M = 0$, $\Psi_e = \psi_1 \psi_2 + \psi_3 \psi_4$ } \\ 
\hline Parton Operators & $\{n_l\}$ & $Q \% 1$ \\ 
\hline 
$\Psi_e$ & 2 0 & $0$ \\ 
$\psi_1 \psi_3 \sim \psi_1 \psi_4 \sim$ ... & 0 2 & $0$ \\ 
$\psi_i$ & 1 1 & $1/2$ \\ 
\hline
\multicolumn{3}{c}{ } \\
\hline
\multicolumn{3}{|c|}{$Z_2$ states, $M = 1$, $\Psi_e = \psi_5(\psi_1 \psi_2 + \psi_3 \psi_4)$} \\
\hline
Parton Operators & $\{n_l\}$ & $Q \% 1$ \\
\hline
$\Psi_e$ & 1 1 0 0 & $0$ \\ 
$\psi_1 \psi_3 \sim \psi_2 \psi_4 \sim$ ... & 0 1 1 0 & $1/2$ \\
$\psi_1 \psi_3 \psi_5 \sim \psi_2 \psi_3 \psi_5 \sim$ ... & 0 0 1 1 & $0$ \\
$\psi_1 \psi_2 \sim \psi_3 \psi_4$ & 1 0 0 1 & $1/2$ \\
$\psi_1 \sim ... \sim \psi_4$ & 1 0 1 0 & $1/4$ \\
$\psi_1 \psi_5 \sim \psi_4 \psi_5 \sim $ ...  & 0 1 0 1 & $3/4$ \\
\hline
\multicolumn{3}{c}{ } \\
\hline
\multicolumn{3}{|c|}{$Z_3$ states, $M = 0$, $\Psi_e = \psi_1 \psi_2 + \psi_3 \psi_4 + \psi_5 \psi_6$} \\
\hline
Parton Operators & $\{n_l\}$ & $Q \% 1$ \\
\hline
$\Psi_e$ & 3 0 & $0$ \\ 
$\psi_i $ & 2 1 & $1/2$ \\
$\psi_1 \psi_3 \sim \psi_1 \psi_4 \sim \psi_1 \psi_5 \sim \psi_1 \psi_6
\sim $ ... & 1 2 & $0$ \\
$\psi_1 \psi_3 \psi_5 \sim \psi_1 \psi_3 \psi_6 \sim$ ... & 0 3 & $1/2$ \\
\hline
\multicolumn{3}{c}{} \\
\hline
\multicolumn{3}{|c|}{$Z_4$ states, $M = 0$, $\Psi_e = \psi_1 \psi_2 + \psi_3 \psi_4 + \psi_5 \psi_6 + \psi_7 \psi_8 $}\\
\hline
Parton Operators & $\{n_l\}$ & $Q \% 1$ \\
\hline
$\Psi_e$ & 4 0 & $0$ \\ 
$\psi_i $ & 3 1 & $1/2$ \\
$\psi_1 \psi_3 \sim \psi_1 \psi_4 \sim \cdots \sim \psi_1 \psi_8\sim$
... & 2 2 & $0$ \\
$\psi_1 \psi_3 \psi_5 \sim \psi_1 \psi_3 \psi_6 \sim $ ... & 1 3 & $1/2$ \\
$\psi_1 \psi_3 \psi_5 \psi_7\sim \psi_1 \psi_3 \psi_6 \psi_8\sim$ ... & 0 4 & $0$ \\
\hline
\end{tabular}
\caption{ \label{pozTable}
We display the pattern of zeros\cite{WW0809,BW0932} 
$\{n_l\}$ for the various parton
operators, and their electromagnetic charge, $Q$,
modulo 1. The operators $\psi_i$ are here chiral free fermion
operators in a 1+1d CFT. Normal ordering is implicit.  There are many
different operators that correspond to topologically equivalent
quasiparticles. Here we listed the ones with minimal scaling
dimension, and $\sim$ indicates gauge equivalences between various
operators. The asymptotic values of the sequence $\{n_l\}$ for large
$l$ classifies each equivalence class. For the $M = 0$ states, each
parton operator $\psi_i$ has electromagnetic charge $q_i = 1/2$. For
the $M = 1$ states, $\psi_i$ has charge $1/4$ for $i = 1,\cdots, 2k$
and $\psi_{2k+1}$ has charge $1/2$.}
\end{table}

The fundamental non-Abelian excitation in the $Z_k$
parafermion states is the excitation that carries minimal charge and
whose fusion with itself can generate all other quasiparticles. In the
projective construction point of view, this operator is $\psi_i$,
for $i = 1, \cdots, 2k$ (they are all gauge-equivalent), and 
corresponds to a single hole in one of the parton IQH states. In the $M = 0$ $Z_k$ parafermion 
states, this operator has electromagnetic charge $Q = 1/2$; its 
scaling dimension can be found using the stress-energy tensor of 
the $U(1)^{2k}/Sp(2k)_1$ theory (see Appendix \ref{lrd}): 
$h_{\psi_i} = 1/2 - (2k+1)/4(k+2) = 3/4(k+2)$, which agrees with the
known results. Notice that for operators with an odd number of parton fields,
the $U(1)^n$ contribution to the scaling dimension is half-integer; this
is related to the fermionic core that we put in by hand when we use the
pure $U(M)\times Sp(2k)$ gauge theory from eqn. (\ref{la1}).

One way to understand how the trivial fermionic holes of the parton
IQH states become non-Abelian excitations is by considering the bulk
effective theory.  The low energy effective theory is a theory of
partons coupled to a $U(M) \times Sp(2k)$ gauge field, which
implements the projection onto the physical Hilbert space. As we
renormalize to low energies, generically a CS term will appear 
for the $U(M) \times Sp(2k)$ gauge field because it is allowed by symmetry. The CS term has
the property that it endows charges with magnetic flux; therefore, two
individual, well-separated partons carry both charge and magnetic flux
in the fundamental representation of $U(M) \times Sp(2k)$. As one
parton is adiabatically carried around another, there will be a
non-Abelian Aharonov-Bohm phase associated with an electric charge
being carried around a magnetic flux. We expect this point of view can be made
more precise in order to compute directly from the bulk theory various
topological properties of the quasiparticles.

\section{Discussion}

We conclude that the correct and most natural description of the
effective field theory for the $Z_k$ parafermion FQH states is the
$U(M) \times Sp(2k)$ CS theory presented here, for which various
topological properties can be explicitly computed. In this case, the
role of the $U(M) \times Sp(2k)$ gauge field is clear: it is to
implement the projection onto the physical Hilbert space generated by
the electron operator.  In particular, the $SU(2)$ quantum numbers are
physical and we should now be able to couple to them through external
probes in the bulk. 

Observe that the electron operator for the $Z_k$ states is a sum of
operators: $\Psi_e = \Psi_{1} + \Psi_{2} + \cdots \Psi_{k}$. This
implies that the $Z_k$ parafermion wave functions can actually be
thought of as a (anti)-symmetrization of a $k$-layer state,
$\Phi_{Z_k} = \mathcal{S}\{\Phi_{abl}(\{z_i^{(l)}\})\}$, where
\begin{align} \Phi_{abl} \sim \langle \prod_{i,l} \Psi_{l}(z^{(l)}_i)
\rangle \end{align} and $z_i^{(l)}$ is the coordinate of the $i$th
electron in the $l$th layer. $\mathcal{S}\{ \cdots \}$ refers to
symmetrization or anti-symmetrization, depending on whether the
particles are boson or fermions, respectively. In the case $M = 0$,
$\Phi_{abl}$ is a $k$-layer wave function with a $\nu = 1/2$ Laughlin
state in each layer. For $M = 1$, it is a generalized $(331)$ wave
function. The fact that the $Z_k$ parafermion wave functions
correspond to (anti)-symmetrizations of these $k$-layer wave functions
was first observed in \Ref{CG0199}. 

The case $k = 2$ corresponds to the Pfaffian, and it is well-known
that the Pfaffian wave function is equal to a symmetrization of the
$(n, n, n-2)$ bilayer wave function, a fact that is closely related to
the existence of a continuous phase transition between the $(n,n,n-2)$
bilayer wave function and the single-layer Pfaffian as the interlayer
tunneling is increased.\cite{RG0067, W0050} These observations suggest
a myriad of possibly continuous phase transitions between various
multilayer Abelian and non-Abelian states as the interlayer tunneling
is tuned, which can be theoretically described by gauge-symmetry
breaking.  For example, breaking the $Sp(2k)$ gauge symmetry down to
$SU(2) \times \cdots \times SU(2)$ would correspond to a phase
transition from a single-layer $Z_k$ parafermion state to a $k$-layer
Abelian state. Breaking $Sp(8)$ to $Sp(4) \times Sp(4)$ could
correspond to a transition between the $Z_4$ parafermion state and a
double layer state with a Pfaffian in each layer. 

Finally, it is interesting to notice that the two ways of thinking about the edge theory and the 
quasiparticle content provide a physical manifestation of the mathematical concept of 
level-rank duality. On the one hand, the edge theory is a projection of free fermions by
the gauge group that keeps the electron operator invariant, while on the other hand, 
it can be understood by considering the representation theory of the electron operator 
algebra. The fact that both perspectives yield the same results is a manifestation of 
level-rank duality. 

This research is supported by NSF Grant No. DMR-0706078.

\appendix

\section{Calculation of Torus Ground State Degeneracy}
\label{gsdeg}

Here we calculate the ground state degeneracy on a torus for the 
$U(1) \times Sp(2k)$ Chern-Simons theory, which is the bulk effective
theory for the $M = 1$ $Z_k$ parafermion states. 
This calculation highlights the fact that simplify specifying the 
gauge group and the level are not enough to fully specify the bulk 
effective theory; one needs also to specify the allowed large gauge 
transformations, which can be done by specifying a choice of electron 
operator. 

For the $M = 1$ $Z_k$ parafermion states, we take the electron operator to be
\begin{equation} 
\label{elOp}
\Psi_e = \psi_{2k+1} \sum_{a=1}^{k} \psi_{2a-1}\psi_{2a}.
\end{equation}
The gauge field takes values in the Lie algebra of $U(1) \times Sp(2k)$,
which in this case consists of $(2k+1)\times (2k+1)$ matrices:
$\left( \begin{matrix} T^a & 0 \\ 0 & 0 \end{matrix} \right)$ and
$diag(0, 1, 0, 1, \cdots, 0, 1, -1)$, with $T^a$ the generators of 
$Sp(2k)$ in the fundamental representation. 

To compute the ground state degeneracy on a torus, we follow the procedure outlined in 
\Ref{W9927}. The classical configuration space of CS theory consists of flat connections,
for which the magnetic field vanishes: $b = \epsilon_{ij} \partial_i a_j = 0$. 
This configuration space is completely characterized by holonomies of the
gauge field along the non-contractible loops of the torus:
\begin{align}
W(\alpha) = \mathcal{P} e^{i \oint_\alpha a \cdot dl}.
\end{align}
More generally, for a manifold $M$, the gauge-inequivalent set of $W(\alpha)$ form
a group: $(\text{Hom: } \pi_1(M) \rightarrow G) /G$, which is the group of
homomorphisms of the fundamental group of $M$ to the gauge group $G$,
modulo $G$. For a torus, $\pi_1(T^2)$ is Abelian, which means that 
$W(\alpha)$ and $W(\beta)$, where $\alpha$ and $\beta$ are the two distinct
non-contractible loops of the torus, commute with each other and we can always 
perform a global gauge transformation so that $W(\alpha)$ and $W(\beta)$
lie in the maximal Abelian subgroup, $G_{abl}$, of $G$ (this subgroup is called the maximal torus).  
The maximal torus is generated by the Cartan subalgebra of the Lie algebra of $G$; in the case at
hand, this Cartan subalgebra is composed of $k + 1$ matrices, $k$ of
which lie in the Cartan subalgebra of $Sp(2k)$, in addition to
$diag(0, 1, 0, 1, \cdots, 0, 1, -1)$. Since we only need to consider
components of the gauge field $a^I$ that lie in the Cartan subalgebra,
the CS Lagrangian becomes 
\begin{align} \mathcal{L} = \frac{1}{4\pi} K_{IJ} a^I \partial a^J,
\end{align}
where $K_{IJ} = \text{Tr}(p^I p^J)$ and $p^I$, $I = 1, \cdots, k+1$ are the 
generators that lie in the Cartan subalgebra. 

There are large gauge transformations $U =  e^{2\pi x_i p^I/L}$, where $x_1$
and $x_2$ are the two coordinates on the torus and $L$ is the length of each
side. These act on the partons as
\begin{equation}
\psi \rightarrow U \psi,
\end{equation}
where $\psi^T = (\psi_1, \cdots, \psi_{2k+1})$, and they take 
$a^I_i \rightarrow a^I_i + 2\pi/L$. These transformations will be the
minimal large gauge transformations if we normalize the generators as follows:
\begin{align}
p^I_{ij} &= \delta_{ij} (\delta_{i,2I} - \delta_{i,2I-1}), \;\;\; I = 1, \cdots, k
\nonumber \\
p^{k+1} &= diag(0, 1, 0, 1, \cdots,0,1, -1).
\end{align}
Thus, for example for the case $k = 3$, the $K$ matrix is
\begin{align}
K = \left(
\begin{matrix}
2 & 0 & 0 & -1 \\
0 & 2 & 0 & -1 \\
0 & 0 & 2 & -1 \\
-1 & -1 & -1 & k+1 \\
\end{matrix} \right)
\end{align}

In addition to the large gauge 
transformations, there are discrete gauge transformations 
$W \in U(1) \times Sp(2k)$ which keep the Abelian subgroup 
unchanged but interchange the $a^I$'s amongst themselves. 
These satisfy
\begin{align}
W^{\dagger} G_{abl} W = G_{abl},
\end{align}
or, alternatively,
\begin{align}
\label{dscTr}
W^{\dagger} p^I W = T_{IJ} p^J,
\end{align}
for some $(k+1) \times (k+1)$ matrix $T$. These discrete transformations correspond 
to the independent ways of interchanging the partons. 

In this $U(1) \times Sp(2k)$ example,
there are $k(k+1)/2$ different discrete gauge transformations $W$. $k$ of them
correspond to interchanging $\psi_{2i-1}$ and $\psi_{2i}$, for $i = 1, \cdots, k$, and
$k(k-1)/2$ correspond to the independent ways of interchanging the $k$ different terms 
in the sum of (\ref{elOp}). 

Picking the gauge $a_0^I = 0$ and parametrizing the gauge field as 
\begin{align}   
a^I_1 = \frac{2\pi}{L}X^I_1 \;\;\;\;\; a^I_2 = \frac{2\pi}{L} X^I_2, 
\end{align}
we have
\begin{align}
L = 2 \pi K_{IJ} X^I_1 \dot X^J_2.
\end{align}
The Hamiltonian vanishes. The conjugate momentum to $X^J_2$ is
\begin{align}
p^J_{2} = 2 \pi K_{IJ} X^I_1.
\end{align}
Since $X_2^J \sim X_2^J + 1$ as a result of the large gauge transformations, 
we can write the wave functions as
\begin{align}
\psi( \vec X_2) = \sum_{\vec n} c_{\vec n} e^{2\pi \vec n \cdot \vec X_2},
\end{align}
where $\vec X_2 = (X_2^1, \cdots X_2^{k+1})$ and $\vec n$ is a $(k+1)$-dimensional vector of integers. 
In momentum space the wave function is
\begin{align}
\phi( \vec p_2 ) &= \sum_{\vec n} c_{\vec n} \delta^{k+1}( \vec p_{2} - 2 \pi \vec n)
\nonumber \\
 &\sim \sum_{\vec n} c_{\vec n} \delta^{k+1}( K \vec X_1 - \vec n),
\end{align}
where $\delta^{k+1}(\vec x)$ is a $(k+1)$-dimensional delta function. 
Since $X_1^J \sim X_1^J + 1$, it follows that $c_{\vec n} = c_{\vec n'}$, where
$(\vec n')^I = n^I + K_{IJ}$, for any $J$. Furthermore, each discrete gauge transformation 
$W_i$ that keeps the Abelian subgroup $G_{abl}$ invariant corresponds to a 
matrix $T_i$ (see eqn. \ref{dscTr}), which acts on the diagonal generators.
These lead to the equivalences $c_{\vec n} = c_{T_i \vec n}$. The number of independent $c_{\vec n}$ can
be computed for each $k$; carrying out the result on a computer, we find
that there are $(k+1)(k+2)/2$ independent wave functions, which agrees with
the known torus ground state degeneracy of the $Z_k$ parafermion states. 

\section{Level-rank duality}
\label{lrd}

To understand the level-rank duality better, let us examine the equivalence
between the $U(1)^{2kn}$ CFT, which is the CFT of $2kn$ free fermions,
and the $Sp(2k)_n \times Sp(2n)_k$ WZW model.  Evidence for the
equivalence of these two theories can be easily established by noting
that they both have the same central charge, $c = 2kn$, and that the
Lie algebra $Sp(2k) \oplus Sp(2n)$ can be embedded into the symmetry
group of the free fermion theory, $O(4kn)$.\cite{FMCFT} The possibility of this
embedding implies that we can construct currents, 
\begin{align} J^A =
\frac{1}{2} \eta_\alpha T^A_{\alpha \beta} \eta_\beta, \;\;\;\;\; J^{a}
= \frac{1}{2} \eta_\alpha T^{a}_{\alpha \beta} \eta_\beta,
\end{align}
where the $\{\eta_\alpha \}$ are Majorana fermions, which are related
to the complex fermions as $\psi_i = \eta_{2i} + i \eta_{2i+1}$.  $\{
T^A \}$ and $\{ T^a \}$ are mutually commuting sets of $4kn \times
4kn$ skew-symmetric matrices that lie in the Lie algebra of $SO(4kn)$
and that separately generate the $Sp(2k)$ and $Sp(2n)$ Lie algebras,
respectively. These currents satisfy the $Sp(2k)_n \times Sp(2n)_k$
current algebra, as can be seen by computing the OPEs: 
\begin{align}
J^A(z) J^B(w) &\sim \frac{n\delta_{AB}}{(z-w)^2} + \frac{i f_{ABC} J^C(w)}{z-w} + \cdots,
\nonumber \\ 
J^{a}(z) J^{b}(w) &\sim \frac{k\delta_{ab}}{(z-w)^2} + \frac{i f_{abc} J^{c}(w)}{z-w} +\cdots,
 \nonumber \\ 
J^{a}(z) J^A(w) &\sim O( (z-w)^0 ).
\end{align}
To compute the levels $n$ and $k$, we have normalized the generators
in the conventional way, so that the quadratic Casimir in the adjoint
representation is twice the dual Coxeter number of the corresponding
Lie algebra. The stress-energy tensor for the $Sp(2k)_n \times
Sp(2n)_k$ theory, defined as 
\begin{align} T(z) = \frac{1}{2 (k + n +
1)} \left( \sum_A J^A J^A + \sum_{a} J^{a} J^{a} \right),
\end{align}
therefore satisfies the same algebra as the stress-energy tensor of the
free fermion theory: $ T_{U(1)}(z) = \frac{1}{2}
\sum_\alpha \eta_\alpha \partial \eta_\alpha$.
Thus, for the $U(1)^{2k}/Sp(2k)_1$ edge theory of the $M = 0$ $Z_k$
parafermion states, we can take the stress tensor to be: \begin{align}
T_{Z_k}(z) &= T_{U(1)}(z) - \frac{1}{2 (k + 2)} \sum_A J^A J^A .
\end{align}
We can use this stress tensor to compute the scaling dimensions of the
quasiparticle operators in the edge theory. 


\end{document}